\documentclass[prl,twocolumn,superscriptaddress]{revtex4}
\usepackage{graphicx}
\usepackage{amsfonts}
\usepackage{amsmath}
\usepackage{amssymb}
\usepackage{bm}
\usepackage{slashed}
\usepackage{dsfont}

\usepackage{color}
\usepackage{soul}

\begin{document}
\title{Emergent Kondo behavior from gauge fluctuations in spin liquids}
\author{Rui Wang}
\email{rwang89@nju.edu.cn}
\affiliation{National Laboratory of Solid State Microstructures and Department of Physics, Nanjing University, Nanjing 210093, China}
\affiliation{Collaborative Innovation Center for Advanced Microstructures, Nanjing 210093, China}
\author{Yilin  Wang}
\affiliation{Hefei National Laboratory for Physical Sciences at Microscale,
University of Science and Technology of China, Hefei, Anhui 230026, China}
\author {Y. X. Zhao}
\affiliation{National Laboratory of Solid State Microstructures and Department of Physics, Nanjing University, Nanjing 210093, China}
\affiliation{Collaborative Innovation Center for Advanced Microstructures, Nanjing 210093, China}
\author{Baigeng Wang}
\email{bgwang@nju.edu.cn}
\affiliation{National Laboratory of Solid State Microstructures and Department of Physics, Nanjing University, Nanjing 210093, China}
\affiliation{Collaborative Innovation Center for Advanced Microstructures, Nanjing 210093, China}

%\date{\today}
\begin{abstract}
Kondo effect is a prominent quantum phenomenon describing the many-body screening of a local magnetic impurity. Here, we reveal a new type of non-magnetic Kondo behavior generated by gauge fluctuations in strongly-correlated baths. We show that a non-magnetic bond defect not only introduces the potential scattering but also locally enhances the gauge fluctuations. The local gauge fluctuations further mediate a pseudospin exchange interaction that produces an asymmetric Kondo fixed point in low-energy.  The gauge-fluctuation-induced Kondo phenomena do not exhibit the characteristic resistivity behavior of conventional Kondo effect, but display a non-monotonous  temperature dependence of thermal conductivity as well as an anisotropic pseudospin correlation. Moreover, with its origin from gauge fluctuations, the Kondo features can be regarded as promising indicators for identifying quantum spin liquids.  Our work advances fundamental knowledge of  novel Kondo phenomena in strongly-correlated systems, which have no counterparts in thermal baths within the single-particle description.
\end{abstract}

\maketitle
\emph{\color{blue}{Introduction.--}}
The Kondo problem, which treats a magnetic impurity
in metals \cite{jkondo}, is of key importance in material science, as
its solution by renormalization group (RG) \cite{kgwilson} and Bethe ansats \cite{nandrei}
invokes some of the most profound concepts and techniques in theoretical physics \cite{achewson}. When a magnetic impurity is coupled to bath electrons, the magnetic scattering becomes essential and drives a many-body resonance. The Kondo singlet is then formed, displaying the Fermi liquid behavior \cite{pnozieres}. In this Letter, we shall extend the scope of Kondo physics to a new avenue, namely describing bond defects in quantum spin liquids (QSLs).

QSLs are exotic states of strongly correlated and frustrated systems in two dimensions, and have constituted one of the most active fields over the last decades~\cite{pwasnders,lsavary}. Due to strong quantum fluctuation, various QSLs  can be stabilized, displaying the fractionalized excitations and emergent gauge field, such as the resonating valence bond (RVB) states \cite{pwandersona} and the flux phases \cite{nread,ehlieb,muubbens,Snir}. The gauge fluctuations are the most crucial degrees of freedom reflecting the nature of QSLs. They can not only ensure the disordered nature of the deconfined mean-field states \cite{xgwenn} but can also generate anyonic excitations via statistical transmutations \cite{Gerald}.

Because of the emergent gauge fluctuations, the impurity problems in QSLs  are complicated, and thus have not been deeply investigated \cite{mogomil,gzchen}. Indeed, there are many important questions to be answered, even if a non-magnetic defect is considered. For example, is there any nontrivial interplay between the defect and the gauge fluctuations? Moreover, will any novel many-body resonances take place, as a result of the gauge fluctuations?

In this Letter, we reveal novel Kondo signatures of a \textit{non-magnetic} defect driven by gauge fluctuations in flux phases.  The flux phases, which describe effective fermions moving under flux \cite{ehlieb}, have been extensively studied in the last decades. They were proposed to enjoy intimate connections with several fundamental topics, including the deconfined quantum criticality \cite{akihiro,cwang}, quantum spin liquids (QSLs) \cite{athomson,xueyangsong}, and high-$T_c$ superconductivity \cite{tchsu}. Here, we unveil salient features of a bond defect in the renowned flux phases. We find that the defect not only introduces the potential scattering but also locally enhances the gauge fluctuations, as shown by Fig.1(a). The gauge fluctuations are able to mediate fermion-fermion interactions.  Here, a pseudospin exchange term between the defect and the bath is locally induced. This leads to a low-energy effective theory formally similar to the Kondo problems in Dirac semimetals or graphene \cite{jiaohaochen,Uchoa,dama,hbzhuang,tffang,linli,akmitchell,ruzheng}, but with several key distinctions. Particularly, the pseudospin exchange interaction is found to be highly anisotropic, and occurs simultaneously with the potential scattering. These distinct features result in an asymmetric Kondo (AK) fixed point. Consequently, a pseudospin Kondo singlet is generated in low-energy with new Kondo features, as shown by Fig.1(b). In sharp contrast with the Kondo effect in normal metals, here we predict a non-monotonous temperature dependence of thermal conductivity as well as an anisotropic pseudospin correlation.

Our work discovers a new mechanism for Kondo behavior in strongly-correlated baths. It implies that the magnetic scattering is no longer necessary for generating Kondo behavior when gauge fluctuations come into play.  More importantly, since the emergent Kondo features are direct consequences of gauge fluctuations, they can serve as promising indicators for QSLs. Our work therefore opens an unprecedented avenue to explore many-body resonances with new mechanisms, which may unveil the mysteries of the emergent gauge fluctuations in QSLs.

\begin{figure}
\includegraphics[width=\linewidth]{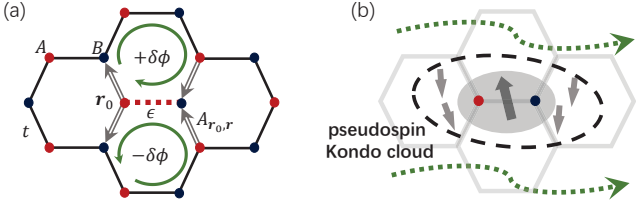}
\caption{Demonstration of the gauge field-induced Kondo physics on honeycomb lattice. (a) A bond defect (the dashed line) with quenched local hopping $t_{\mathbf{r}_0A,\mathbf{r}_0B}=\epsilon<t$ is considered on top of the flux phase, where $t$ is the nearest neighbor hopping.  For $\epsilon\rightarrow 0$, the two hexagons adjacent to the quenched bond become connected, forming a doubled plaquette.  Opposite fluctuations, $+\delta \phi$ and $-\delta \phi$, are enabled in the two hexagons, which can be equivalently represented by the local gauge terms, $A_{\mathbf{r}_0,\mathbf{r}}$,  on the four bonds denoted by the doubled lines.  (b) The defect consists of two sites with A and B sublattice (the red and blue dot), forming an effective local pseudospin moment (the thick arrow in the shaded region). The bath fermions then screen the effective local momentum, resulting in a pseudospin Kondo singlet.}
\end{figure}

\textit{\color{blue}{Chern-Simons representation of flux phases.--}}  We firstly derive the flux phases from spin-1/2 XXZ model based on the Chern-Simons (CS) representation  \cite{alopez,kyang,ruia,tsedrakyana,tsedrak,ttased,ruib}, which will facilitate the study of  impurity problem.

The spin-raising and lowering operators are statistically equivalent to hardcore bosons. To avoid the hardcore condition, we represent the spin operators by spinless fermions. Moreover, in order to ensure the bosonic statistics,  a flux quanta has to be attached to each fermion, as shown by Fig.2(a). This constitutes an exact representation of the spin excitations \cite{sup}.

The attachment of a flux quanta can be achieved by coupling a CS gauge field $A_{\mu}$ \cite{alopez} to the fermions, with the action
\begin{equation}\label{eq3}
S_{CS}=\frac{1}{4\pi}\int d^2xdt \epsilon^{\mu\nu\rho}A_{\mu}\partial_{\nu}A_{\rho},
\end{equation}
which is a pure gauge theory that has no contribution to the system energy \cite{Gerald}.

We now consider a fermion hopping from  $\mathbf{r}$ to $\mathbf{r}^{\prime}$ on the honeycomb lattice as an example. An additional phase will be generated during the hopping, namely,
$A_{\mathbf{r},\mathbf{r}^{\prime}}=\int^{\mathbf{r}^{\prime}}_{\mathbf{r}}d\mathbf{r}^{\prime\prime}\cdot\mathbf{A}(\mathbf{r}^{\prime\prime})$, where $\mathbf{A}(\mathbf{r})$ is the spatial component of $A_{\mu}$. Moreover, the flux in any plaquette can be rigorously obtained by the contour integral as  \cite{sup}
\begin{equation}\label{eq0}
\phi_{\mathbf{r}}=\oint d\mathbf{r}^{\prime\prime}\cdot\mathbf{A}(\mathbf{r}^{\prime\prime})=2\pi n_{\mathbf{r}}.
\end{equation}
Hereby, we have labelled the plaquette by its bottom-left site $\mathbf{r}$. Eq.\eqref{eq0} constitutes a flux condition in the CS fermion representation, which requires that the flux in the $\mathbf{r}$ plaquette be proportional to the local fermion number.
 \begin{figure}
\includegraphics[width=\linewidth]{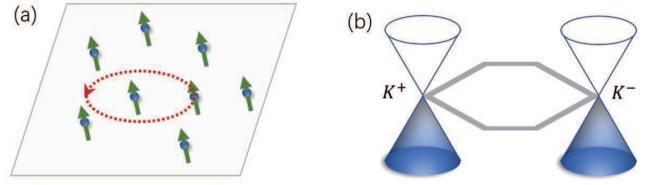}
\caption{The flux attachment and emergent flux phase. (a) The CS representation of the spin excitations. Each spinless fermion is attached to a flux quanta, reproducing the $\mathrm{SU}$(2) algebra of the spin operators.  (b) On honeycomb lattice, the saddle point solution is obtained, i.e., the flux phase with $2\pi$ flux in each hexagon. This state exhibits  low-energy Dirac fermions with the valley degrees of freedom, $K^+$ and $K^{-}$.  }
\end{figure}

Using the above formalism, the XXZ model can be fermionized as $H=H_0+H_{int}$ \cite{sup}, where
\begin{equation}\label{eq1}
\begin{split}
  H_0&=\sum_{\mathbf{r},\mathbf{r}^{\prime}}(t_{\mathbf{r},\mathbf{r}^{\prime}}f^{\dagger}_{\mathbf{r}}e^{iA_{\mathbf{r},\mathbf{r}^{\prime}}}f_{\mathbf{r}^{\prime}}+h.c.),
\end{split}
\end{equation}
comes from the XY term, and $H_{int}=\sum_{\mathbf{r},\mathbf{r}^{\prime}}u_{\mathbf{r},\mathbf{r}^{\prime}}(n_{\mathbf{r}}-1/2)(n_{\mathbf{r}^{\prime}}-1/2)$ is the local interaction arising from the Ising term. We note that $H$ has particle-hole symmetry.

The gauge field in Eq.\eqref{eq3} and Eq.\eqref{eq1} has fluctuations.  However, for stable flux phases, the gauge fluctuations are irrelevant \cite{michaelhermele} in the sense that they only provide a flux background that modulates the energy of fermions \cite{sup,michaelhermele}. Then, Eq.\eqref{eq1} is reduced to the exact flux model investigated by E. Lieb \cite{ehlieb}. It was proved that the saddle point of Eq.\eqref{eq1} corresponds to the  $2\pi$-flux phase \cite{ehlieb}, where $\phi_{\mathbf{r}}=2\pi$ (mod $2\pi$)  in all the hexagons. Moreover, $H_{int}$ is found irrelevant for $u_{\mathbf{r},\mathbf{r}^{\prime}}\lesssim t_{\mathbf{r},\mathbf{r}^{\prime}}$.  Correspondingly, as shown by Fig.2(b), two low-energy Dirac fermions emerge, located at $\mathbf{K}^a$ in momentum space, with the valley index $a=\pm$, i.e.,
\begin{equation}\label{eq2}
  H_{0}=v_F\int\frac{d^2k}{(2\pi)^2}f^{(a)\dagger}_{\mathbf{k},\alpha}\boldsymbol{\sigma}^{(a)}_{\alpha\beta}\cdot\mathbf{k}f^{(a)}_{\mathbf{k},\beta},
\end{equation}
where sum of repeated indices is understood. $\boldsymbol{\sigma}^{(a)}$ is the Pauli matrix defined in the pseudospin (sublattice) space and is generally valley-dependent. Eq.\eqref{eq2} together with Eq.\eqref{eq3} concisely describes the flux phases under the CS representation, as long as they are stabilized.

\textit{\color{blue}{Local gauge fluctuations excited by defect.}}--We now consider the bond defect, with the hopping  $\epsilon<t$ and $t$ the nearest neighbor (n.n.) hopping.  The Hamiltonian in Eq.\eqref{eq1} still respects the particle-hole symmetry, as well as a reflection symmetry with respect to the bond center,  $\mathbf{P}$, as shown by  Fig.3(a).  In this case, the Lieb's theorem is applicable, which states that the $2\pi$ flux is maintained in the plaquettes that intersect $\mathbf{P}$ \cite{ehlieb}.
%%%%%%%%%

So far, we neglected the gauge fluctuations, which are irrelevant for flux phases stabilized on perfect lattices. Since a defect is now involved, the gauge fluctuations need to be considered more carefully.  As indicated by Fig.1(a), for $\epsilon\rightarrow0$, the two  hexagons sharing the bond become connected, forming a doubled plaquette.  Then, opposite fluctuations, $+\delta\phi$ and $-\delta\phi$, can naturally emerge in each of the two hexagons. Thus,  a sufficiently strong bond defect with $\epsilon\ll t$  is expected to locally enhance the gauge fluctuations.

To verify the above expectation, we perform a self-consistent calculation on a finite lattice \cite{sup}, with the fluxes treated as variational parameters. We start from initial states with randomly-generated fluxes and optimize them by minimizing the system energy. Fast convergence to $2\pi$ flux is observed for most plaquettes. However, for $\epsilon\ll t$, it is found that the fluxes at the defect site, $\delta\phi_{\mathbf{r}_0}$, can hardly converge, and they display a dependence on the initial states. Such numerical fluctuations essentially reflect the strong gauge fluctuations around the saddle point. To describe the fluctuation, we define  $\overline{\phi}_{\mathbf{r}}=\langle \phi_{\mathbf{r}}\rangle$ and  $\delta \phi_{\mathbf{r}}=(\langle \phi^2_{\mathbf{r}}\rangle-\langle \phi_{\mathbf{r}}\rangle ^2)^{1/2}$, where $\langle...\rangle$ denotes the expectation over the random initial states. Clearly, $\overline{\phi}_{\mathbf{r}}$ and $\delta \phi_{\mathbf{r}}$ are numerical simulations of the saddle point fluxes and the gauge fluctuations, respectively.

\begin{figure}
\includegraphics[width=\linewidth]{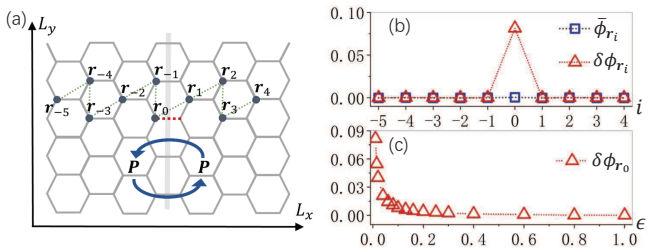}
\caption{The effect of the bond defect on the flux phase.   (a) The quenched bond is placed at the center of a $8\times10$ lattice. The system respects the reflection symmetry $\mathbf{P}$. (b) $\overline{\phi}_{\mathbf{r}}$ and  $\delta\phi_{\mathbf{r}}$ obtained by the gradient descendant method \cite{sup}. The expectations are evaluated with 200 random initial flux configurations. The x-axis denotes the 10 plaquettes along the dashed trajectory in (a), from the $\mathbf{r}_{-5}$ to $\mathbf{r}_4$. The $\mathbf{r}_0$-plaquette is the one adjacent to the quenched bond, whose hopping is taken as $\epsilon=0.01$. (c) $\delta \phi_{\mathbf{r}_0}$ as a function of  $\epsilon$. The n.n. hopping is taken as $t=1$.}
\end{figure}

We calculate $\overline{\phi}_{\mathbf{r}}$ and $\delta \phi_{\mathbf{r}}$ along the dashed trajectory in Fig.3(a). As shown in Fig.3(b), $\overline{\phi}_{\mathbf{r}}$ converges to $2\pi$ (gauge equivalent to 0) for all the plaquettes.  Moreover,  although $\delta \phi_{\mathbf{r}}=0$ is obtained for most plaquettes, $\delta \phi_{\mathbf{r}_0}$ exhibits a significant fluctuation. In addition, as shown by Fig.3(c), $\delta \phi_{\mathbf{r}_0}$ firstly changes slowly with decreasing $\epsilon$, and then grows fast for $\epsilon\ll t$, implying that  only a strong defect can enhance the local gauge fluctuations.

The above numerical results reveal two important facts. First, the saddle point of Eq.\eqref{eq1} is still the $2\pi$-flux phase, even if a bond defect is taken into account, in accordance with Lieb's theorem \cite{ehlieb}. Second, the defect further enables local gauge fluctuations around the saddle point, in the plaquettes nearby the defect, in consistent with  the intuitive picture illustrated by Fig.1(a).

Let us now focus on the defect site $\mathbf{r}_0$.  The $\mathbf{r}_0$ plaquette has the saddle point flux $2\pi$ (mod $2\pi$),  as was proved by both the Lieb's theorem and the self-consistent calculations. Moreover, we recall that the flux must be bound to the fermion number according to Eq.\eqref{eq0}. Therefore, a local constraint at $\mathbf{r}_0$ is obtained, namely, $n_{\mathbf{r}_0}=n_{\mathbf{r}_0,A}+n_{\mathbf{r}_0,B}=0,1,2$.  Then, the particle-hole symmetry further fixes  a single-occupation constraint condition \cite{sup},  $n_{\mathbf{r}_0,A}+n_{\mathbf{r}_0,B}=1$. Thus, the site $\mathbf{r}_0$ can only be occupied by a single fermion with pseudospin A or B.  An effective impurity therefore emerges, displaying a pseudospin-1/2 moment, as indicated by Fig.1(b). In order to facilitate the following analysis and distinguish the impurity fermions from the rest,  we introduce the notation $d_{\mathbf{r}_0,\alpha}=f_{\mathbf{r}_0,\overline{\alpha}}$. Then, the single-occupation condition is cast into the simple form
\begin{equation}\label{eq4}
  \sum_{\alpha}d^{\dagger}_{\mathbf{r}_0,\alpha}d_{\mathbf{r}_0,\alpha}=1.
 \end{equation}
 In addition to Eq.\eqref{eq4}, we have also shown that the fluctuation $\delta\phi_{\mathbf{r}_0}$ around the saddle point has to be taken into account for $\epsilon\ll t$.  This can be equivalently written into local hoppings between the impurity states ($d_{\mathbf{r}_0,\alpha}$) and the bath fermions ($f_{\mathbf{r},\alpha}$), i.e.,
\begin{equation}\label{eq5}
H^{loc}=\sum_{\mathbf{r},\alpha}t_{\mathbf{r},\mathbf{r}_0}f^{\dagger}_{\mathbf{r},\alpha}e^{iA_{\mathbf{r},\mathbf{r}_0}}d_{\mathbf{r}_0,\alpha}+h.c.,
\end{equation}
where the CS gauge field $A_{\mathbf{r},\mathbf{r}_0}$ participates in the hopping processes. The sum here involves four local terms, as indicated by the double-lined bonds in Fig.1(a).

 \textit{\color{blue}{Gauge field-induced pseudospin Kondo model.}}--  So far we have rigorously derived an effective model for $\epsilon\rightarrow0$, which consists of a pseudospin-1/2 impurity  ($d_{\mathbf{r}_0,\alpha}$) and a thermal bath ($f_{\mathbf{r},\alpha}$).  The bath, in essence, is a $2\pi$-flux phase with two vacancies at $\mathbf{r}_0$. It is thus described by
\begin{equation}\label{eq6}
H^{bath}=v_F\int\frac{d^2k}{(2\pi)^2}f^{(a)\dagger}_{\mathbf{k},\alpha}\boldsymbol{\sigma}^{(a)}_{\alpha\beta}\cdot\mathbf{k}f^{(a)}_{\mathbf{k},\beta}+Vf^{\dagger}_{\mathbf{r}_0,\alpha}f_{\mathbf{r}_0,\alpha},
\end{equation}
where the strong local potential $V$ efficiently removes the two sites at $\mathbf{r}_0$ \cite{Pereira}.

As discussed, the bath fermions are coupled to the effective impurity via Eq.\eqref{eq5}, which includes the local CS gauge field, $A_{\mathbf{r},\mathbf{r}_0}$.  It is well known that the gauge field can mediate the electron-electron interaction in quantum electrodynamics. By analogy, here, the local CS gauge field, $A_{\mathbf{r},\mathbf{r}_0}$, will induce a local interaction between the bath and impurity. We therefore integrate out the local gauge fluctuations in Eq.\eqref{eq3} and Eq.\eqref{eq5}. In the continuum limit, a four-fermion term is generated as    \cite{sup}
\begin{equation}\label{eq7}
\delta H=J_{eff}\sum_{\alpha}f^{\dagger}_{\mathbf{r}_0,\alpha}f_{\mathbf{r}_0,\overline{\alpha}}d^{\dagger}_{\mathbf{r}_0,\overline{\alpha}}d_{\mathbf{r}_0,\alpha}.
\end{equation}
Clearly, this describes the pseudospin-flip processes when the bath fermions ($f_{\mathbf{r},\alpha}$) are scattered by the impurity ($d_{\mathbf{r}_0,\alpha}$).  The gauge fluctuations here act as a ``strong glue", and the resultant coupling constant $J_{eff}$ is found to dominate over the energy scale of the Dirac fermions described by $H^{bath}$.

We observe a great similarity between Eq.\eqref{eq7} and the $\mathrm{SU}(2)$ Coqblin-Schrieffer model \cite{bcoqblin}. Indeed, after introducing the $\mathrm{SU}(2)$ generator,  $X_{{\alpha\beta}}=d^{\dagger}_{\alpha}d_{\beta}-\delta_{\alpha\beta}/2$, and $X_{12}\sim S^+$, $X_{21}\sim S^-$, Eq.\eqref{eq7} along with the single-occupation condition in Eq.\eqref{eq4} is exactly mapped to the pseudospin exchange interaction,
\begin{equation}\label{eq9}
H_K=J_{eff}\mathbf{S}_p\cdot(f^{\dagger}_{\mathbf{r}_0,\alpha} \boldsymbol{\sigma}_{\alpha\beta}f_{\mathbf{r}_0,\beta}),
\end{equation}
where $\mathbf{S}_p=(S_x,S_y)$ is the effective pseudospin-1/2 operator at $\mathbf{r}_0$. Since only $(S_x,S_y)$ is present, the interaction $H_K$ is highly anisotropic.

We have finally arrived at the low-energy effective theory for the renowned flux phases with a strong bond defect, i.e., Eq.\eqref{eq6} and Eq.\eqref{eq9}. Interestingly, an anisotropic pseudospin Kondo model emerges on top of the Dirac fermions with two valleys and a strong local potential $V$.

  \textit{\color{blue}{Asymmetric Kondo fixed point.}}--Eq.\eqref{eq6} and Eq.\eqref{eq9} are similar in their form to the Kondo problems in Dirac semimetals or graphene \cite{jiaohaochen,Uchoa,dama,hbzhuang,tffang,linli,akmitchell}. However, there are several key distinctions. First,  the bath is composed of effective fermions deconfined from  the original quantum spin model. They neither carry charge nor spin, but pseudospin.  Second, the bath fermions exhibit a pseudospin-momentum locking, as explicit in Eq.\eqref{eq6}. Their pseudospins are therefore fully polarized.  Third, the exchange interaction in Eq.\eqref{eq9} is not $\mathrm{SU}(2)$-invariant, but highly anisotropic with only the XY coupling.  Last, the potential scattering $V$ and the pseudospin exchange $J_{eff}$ simultaneously take place in Eq.\eqref{eq6} and Eq.\eqref{eq9}. Thus,  the interplay between them is non-negligible \cite{akmitchell}.

We firstly perform a RG analysis, treating $V$ and $J_{eff}$ as perturbations \cite{sup}. To two-loop order, the RG flow is obtained as $dJ_{eff}/dl=- J_{eff}+J^2_{eff}-J^3_{eff}/2$ and $dV/dl=-V$. Both $V$ and $J_{eff}$ are irrelevant, flowing to the local momentum (LM) fixed point with $(V,J_{eff})=(0,0)$, as shown by the blue dot in Fig.4(a).

The above perturbative RG flow breaks down for large $V$ and $J_{eff}$, therefore not applicable to the current case. For large $V$ and $J_{eff}$, we identify a stable AK fixed  \cite{Carlos,larsfritz} with $(V,J_{eff})=(\infty,\infty)$, based on the duality between the strong-coupling Kondo model and the  weak-coupling Anderson model \cite{achewson,qlli}. By investigating the RG flow of the dual Anderson model, the fixed points far away from $(V,J_{eff})=(0,0)$ can be derived \cite{sup}, producing the complete RG flow as shown by  Fig.4(a).  Interestingly, since this emergent Kondo problem is characterized by large $V$ and $J_{eff}$, the system is always located in the ``basin of attraction" of the AK fixed point, as indicated by the shaded regime in Fig.4(a).
\begin{figure}
\includegraphics[width=\linewidth]{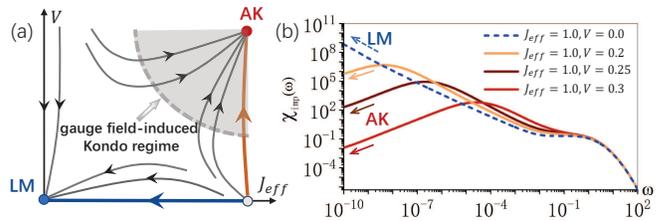}
\caption{The fixed points of the effective theory. (a) The LM and AK fixed point are found, together with a third unstable fixed point with $(V,J_{eff})=(0,\infty)$. The gauge-field assisted Kondo model lies in a basin of attraction governed by the AK fixed point, as indicated by the shaded regime. (b) The NRG results of the imaginary part of the dynamical spin susceptibility $\chi_{imp}(\omega)$. The system always flows to the LM fixed point for $V=0$, corresponding to the horizontal blue RG trajectory in (a). The Fermi liquid behavior with $\chi_{imp}(\omega)\propto\omega$ is found for finite V and large $J_{eff}$, justifying the flow to the AK fixed point in correspondence to the thick orange trajectory in (a).}
\end{figure}

The AK fixed point naturally arises as a result of the interplay between $V$ and $J_{eff}$. For $V=0$, $J_{eff}$ is always irrelevant due to the vanishing density of states (DOS) at the Dirac point. The scattering $V$, in a sense, acts as a local chemical potential and enhances the local DOS (LDOS), which in turn favors the Kondo screening.
We further perform a full-density matrix (FDM) NRG calculation \cite{sup,Andreasa,Andreasb}. Fig.4(b) shows the calculated dynamical spin susceptibility $\chi_{imp}(\omega)$ for different $V$ with $J_{eff}=1$.  In low-energy, the calculated $\chi_{imp}(\omega)$ displays the scaling $\chi_{imp}(\omega)\propto\omega$ for $V\neq0$, implying the occurrence of the Fermi liquid behavior from the Kondo fixed point. Therefore, the system flows to the AK  (LM) fixed point for $V\neq0$ ($V=0$),  in consistent with the thick orange (blue) RG trajectory in Fig.4(a).

So far, all the results are obtained for a short-range coupling, $J_{eff}$. If one assumes a long-range $|\mathbf{r}|$-dependence of $J_{eff}(|\mathbf{r}|)$, then the inter-valley scattering amplitude will become smaller compared to the  intra-valley one. In this case, the two-channel Kondo physics can be relevant \cite{Affleckb,sup}, resulting in the non-Fermi liquid fixed point as well as the crossover between the Fermi liquid and non-Fermi liquid behavior.

\textit{\color{blue}{Conclusion and discussion.}}--   The predicted Kondo behavior exhibits several unique features that are experimentally measurable. First, although the local Kondo resonance arises from an unusual mechanism,  it will still exhibit a LDOS peak near the Fermi energy.  The STM/STS measurements may be applicable for measuring the impurity LDOS. Second, since the screening is formed by deconfined and chargeless excitations rather than electrons, the characteristic resistivity behavior of conventional Kondo effect does not take place here. Third, the nature of fixed point can be manifested by thermal transport \cite{Annamaria,moca}.  We have calculated the thermal conductivity $\sigma_E(T)/T$  at low energy scales near the AK fixed point based on the conformal field theory \cite{Affleckb,Afflecka,Affleckc,edwitten,Ludwig,Affleckd,Afflecke,Ludwigf}. It is found that $\sigma_E(T)/T$ scales as $\sigma_E(T)/T\sim T^2$ at low temperatures  \cite{sup}. Moreover, it is known that the potential scattering $V$ can bring about significant deviations from the universal scaling at higher temperatures ($T\gtrsim T_K$, with $T_K$ the Kondo temperature) \cite{jkondob,Annamaria}. Therefore, a non-monotonous temperature dependence of the thermal conductivity is expected. Last, because of the pseudospin-momentum locking of the bath, both the pseudospin and the orbital angular momentum will participate in the screening process. This always leads to an anisotropic pseudospin correlation, which has been proved by Ref.\cite{ruiwangc} in the context of topological superconductors.

This work reveals a gauge-fluctuation-induced Kondo behavior. The Kondo screening here is caused by gauge fluctuations, and therefore has a fundamentally different origin from the conventional Kondo effect. Nevertheless, it provides  a promising indicator for identifying QSLs.
These results are generalizable to flux phases stabilized on different lattices and other physical systems.  \cite{sup}. Particularly, the fermions coupled to stable gauge fluxes can also emerge from interacting bosonic systems with degenerate single-particle dispersions \cite{tigranadd,tigranaddb,Saurabh}, e.g., the dilute atomic gases \cite{jamesr}.
Furthermore, it is highly desirable to verify our theory by large-scale numerical simulations. For example, the quantum Monte Carlo calculations may be applied if the sign problem can be avoided in some flux phases \cite{xiaoyan}.

\begin{acknowledgments}
We thank Andreas Weichselbaum (Brookhaven National Laboratory, USA) and Seung-Sup Lee (Ludwig Maximilian University of Munich, Germany) for providing us the QSpace tensor library and the NRG code for arbitrary type of bath, respectively. R. W. is  grateful to Tigran Sedrakyan, Xiaoqun Wang, Qianghua Wang, Peng Song and W. Su. for fruitful discussions.   This work was supported by the Youth Program of National Natural Science Foundation of China (No. 11904225) and  the National Key  R\&D Program of China (Grant No. 2017YFA0303200). Y. W. was supported by USTC Research Funds of the Double First-Class Initiative (No. YD2340002005)
\end{acknowledgments}

%%%%%%%%%% Merge with supplemental materials %%%%%%%%%%
\pagebreak
\vspace{5cm}
\widetext
%%%%%%%%%% Merge with supplemental materials %%%%%%%%%%
%%%%%%%%%% Prefix a "S" to all equations, figures, tables and reset the counter %%%%%%%%%%
\setcounter{equation}{0}
\setcounter{figure}{0}
\setcounter{table}{0}
\setcounter{page}{1}
\makeatletter
\renewcommand{\theequation}{S\arabic{equation}}
\renewcommand{\thefigure}{S\arabic{figure}}
\renewcommand{\bibnumfmt}[1]{[S#1]}
\renewcommand{\citenumfont}[1]{S#1}
%%%%%%%%%% Prefix a "S" to all equations, figures, tables and reset the counter %%%%%%%%%%

\pagebreak
\vspace{5cm}
\widetext
\begin{center}
\textbf{\large Supplemental material for: Emergent Kondo behavior from gauge fluctuations in spin liquids}
\end{center}

%Compiling Mode: pdflatex
%\usepackage{graphicx}
%\usepackage{amsfonts}
%\usepackage{amsmath}
%\usepackage{amssymb}
%\usepackage{bm}
%\usepackage{hyperref}
%\usepackage{slashed}

%\title{Supplementary material for ``Antiferromagnetic magnons from fractionalized excitations"}

%\author{Baigeng Wang$^{2}$}
%\author{L. Sheng$^{2}$}
%\author{D. Y. Xing$^{2}$}
%\author{Jian Wang$^{1,3}$}

\date{\today }
\maketitle
\section{the flux phases under Chern-Simons representation}
In this section, we assume that the flux phases  are stabilized and derive them from the quantum  XXZ model, using the Chern-Simons representation.  We start from the XXZ model on hexagonal lattice as follows,
\begin{equation}\label{eqs1}
H_{XXZ}=\sum_{\mathbf{r},\mathbf{r}^{\prime}}J_{\mathbf{r},\mathbf{r}^{\prime}}(S^x_{\mathbf{r}}S^x_{\mathbf{r}^{\prime}}+S^y_{\mathbf{r}}S^y_{\mathbf{r}^{\prime}})+\sum_{\mathbf{r},\mathbf{r}^{\prime}}J^z_{\mathbf{r},\mathbf{r}^{\prime}}S^z_{\mathbf{r}}S^z_{\mathbf{r}^{\prime}}.
\end{equation}
We fermionize the spin operator as spinless fermions coupled to CS string operators as
\begin{equation}\label{eqs2}
  S^{+}_{\mathbf{r}}=f^{\dagger}_{\mathbf{r}}\alpha^+_{\mathbf{r}}, ~ S^{}_{\mathbf{r}}=f_{\mathbf{r}}\alpha^-_{\mathbf{r}},
\end{equation}
where $\alpha^{\pm}_{\mathbf{r}}=\sum_{\mathbf{r}^{\prime}\neq\mathbf{r}}\mathrm{arg}(\mathbf{r}-\mathbf{r}^{\prime})n_{\mathbf{r}^{\prime}}$ are the string operators.
After fermionization, one obtains,
\begin{equation}\label{eqs3}
H_{XXZ}=\sum_{\mathbf{r},\mathbf{r}^{\prime}}[t_{\mathbf{r},\mathbf{r}^{\prime}}f^{\dagger}_{\mathbf{r}}e^{i(A_{\langle\mathbf{r},\mathbf{r}^{\prime}\rangle}+\delta A_{\mathbf{r},\mathbf{r}^{\prime}})}f_{\mathbf{r}^{\prime}}+h.c.]+\sum_{\mathbf{r},\mathbf{r}^{\prime}}u_{\mathbf{r},\mathbf{r}^{\prime}}(n_{\mathbf{r}}-1/2)(n_{\mathbf{r}^{\prime}}-1/2),
\end{equation}
where $A_{\langle\mathbf{r},\mathbf{r}^{\prime}\rangle}$ denotes the  non-dynamical CS phases, and $\delta A_{\mathbf{r},\mathbf{r}^{\prime}}$ describes the fluctuation of the CS gauge field, with a CS action $S_{CS}$. As discussed in the main text, for stable flux phases, one can neglect the gauge fluctuations, $\delta A_{\mathbf{r},\mathbf{r}^{\prime}}$.  Then, the fermionized XXZ Hamiltonian is reduced to the model studied by E. Lieb. Moreover, a self-consistent calculation can be performed to determine $A_{\langle\mathbf{r},\mathbf{r}^{\prime}\rangle}$.  Either based on Lieb's proof or the self-consistent calculation, one can unveil the saddle point solution of Eq.\eqref{eqs3}, which is found to be the $2\pi$-flux phase. Gauging out the $2\pi$-flux in each plaquette, Eq.\eqref{eqs3} is simplified to
\begin{equation}\label{eqs4}
H_{XXZ}=\sum_{\mathbf{r},\mathbf{r}^{\prime}}[t_{\mathbf{r},\mathbf{r}^{\prime}}f^{\dagger}_{\mathbf{r}}f_{\mathbf{r}^{\prime}}+h.c.]+\sum_{\mathbf{r},\mathbf{r}^{\prime}}u_{\mathbf{r},\mathbf{r}^{\prime}}(n_{\mathbf{r}}-1/2)(n_{\mathbf{r}^{\prime}}-1/2).
\end{equation}
The first term describes the Dirac CS fermions in low-energy, while the second describes the local interaction between fermions. Such interaction is irrelevant for 2D Dirac fermions in renormalization group sense.

\section{the pseudospin exchange interaction induced by local gauge fluctuation}
In main text, we have derived the total Hamiltonian, $H^{tot}=H^{bath}+H^{loc}$, where $H^{bath}$ is the $2\pi$-flux phase but with two sites at the defect being removed, described by Eq.(7) of the main text. The local gauge fluctuation term $H^{loc}$, given by Eq.(6) of the main text, is expanded into
\begin{equation}\label{eqs5}
H_{loc}=t(f^{\dagger}_{\mathbf{r}_0+\mathbf{e}_1,B}e^{iA_{\mathbf{r}_0+\mathbf{e}_1,\mathbf{r}_0}}d_{\mathbf{r}_0,B}+f^{\dagger}_{\mathbf{r}_0-\mathbf{e}_1,A}e^{iA_{\mathbf{r}_0-\mathbf{e}_1,\mathbf{r}_0}}d_{\mathbf{r}_0,A}+f^{\dagger}_{\mathbf{r}_0+\mathbf{e}_3,B}e^{iA_{\mathbf{r}_0+\mathbf{e}_3,\mathbf{r}_0}}d_{\mathbf{r}_0,B}+f^{\dagger}_{\mathbf{r}_0-\mathbf{e}_3,A}e^{iA_{\mathbf{r}_0-\mathbf{e}_3,\mathbf{r}_0}}d_{\mathbf{r}_0,A})+h.c.
\end{equation}
where  $\mathbf{e}_i$ ($i=1,2,3$) are the three nearest neighbor (n.n.) bond vectors on honeycomb lattice.  Neglecting the gauge fluctuation and in the continuum limit where $\mathbf{e}_i\rightarrow0$, Eq.\eqref{eqs5} reduces to simple hopping terms between bath fermions and the impurity states, i.e.,
\begin{equation}\label{eqs6}
H^{loc}_{0}=2t\sum_{\alpha}(f^{\dagger}_{\mathbf{r}_0,\alpha}d_{\mathbf{r}_0,\alpha}+h.c.)
\end{equation}
However, as shown by the main text, although the flux phase is still the saddle point solution, the local gauge fluctuation is nevertheless enhanced by the defect. Furthermore, for a sufficiently strong defect with $\epsilon\sim0.01$, $\delta\phi_{\mathbf{r}_0}\sim0.08$ is obtained, as in Fig.3 of the main text, therefore the gauge phases $A_{\mathbf{r},\mathbf{r}_0}\sim \delta\phi_{\mathbf{r}_0}/2\sim0.04\ll1$ in the Eq.\eqref{eqs5}. This therefore enables the expansion with respect to $A$'s. Besides the zeroth order term, $H^{loc}_0$, the first order expansion leads to
\begin{equation}\label{eqs7}
H^{loc}_{a}=it(f^{\dagger}_{\mathbf{r}_0+\mathbf{e}_1,B}A_{\mathbf{r}_0+\mathbf{e}_1,\mathbf{r}_0}d_{\mathbf{r}_0,B}+f^{\dagger}_{\mathbf{r}_0-\mathbf{e}_1,A}A_{\mathbf{r}_0-\mathbf{e}_1,\mathbf{r}_0}d_{\mathbf{r}_0,A}+f^{\dagger}_{\mathbf{r}_0+\mathbf{e}_3,B}A_{\mathbf{r}_0+\mathbf{e}_3,\mathbf{r}_0}d_{\mathbf{r}_0,B}+f^{\dagger}_{\mathbf{r}_0-\mathbf{e}_3,A}A_{\mathbf{r}_0-\mathbf{e}_3,\mathbf{r}_0}d_{\mathbf{r}_0,A})+h.c.
\end{equation}
Then, the total Hamiltonian reads as $H^{tot}=H^{bath}+H^{loc}_{0}+H^{loc}_{a}$, where the gauge degrees of freedom enjoys the CS action $S_{CS}$. The zero-temperature partition function then can be written as
 \begin{equation}\label{eqs8}
Z=\int D\overline{f}DfD\overline{d}Dd e^{i(S^{bath}+S^{loc}_{0})}\int DA_{\mu}e^{i(S^{loc}_{a}+S_{CS})},
 \end{equation}
where the action $S^{bath}$ and $S^{loc}_0$ are independent of gauge field $A_{\mu}$, therefore are separated from the functional integral of $A_{\mu}$. The $A_{\mu}$-dependent terms, $S^a_{loc}$ and $S_{CS}$ are respectively given by
\begin{equation}\label{eqs9}
S^{loc}_a=it\int dx_0(f^{\dagger}_{\mathbf{r}_0+\mathbf{e}_1,B}A_{\mathbf{r}_0+\mathbf{e}_1,\mathbf{r}_0}d_{\mathbf{r}_0,B}+f^{\dagger}_{\mathbf{r}_0-\mathbf{e}_1,A}A_{\mathbf{r}_0-\mathbf{e}_1,\mathbf{r}_0}d_{\mathbf{r}_0,A}+f^{\dagger}_{\mathbf{r}_0+\mathbf{e}_3,B}A_{\mathbf{r}_0+\mathbf{e}_3,\mathbf{r}_0}d_{\mathbf{r}_0,B}+f^{\dagger}_{\mathbf{r}_0-\mathbf{e}_3,A}A_{\mathbf{r}_0-\mathbf{e}_3,\mathbf{r}_0}d_{\mathbf{r}_0,A})+h.c.
\end{equation}
and
\begin{equation}\label{eqs10}
S_{SC}=\int \frac{d^2qd\Omega}{(2\pi)^3}A_{\mu,\mathbf{q}}(\gamma q^{\mu}q^{\nu}-\frac{i}{4\pi}\epsilon^{\mu\nu\lambda}q_{\lambda})A_{\nu,-\mathbf{q}},
\end{equation}
where $\gamma$ is a the gauge fixing factor that takes $\gamma\rightarrow\infty$ in later steps. Then, the propagator of CS gauge field is obtained as
\begin{equation}\label{eqs11}
D^{\mu\nu}_{\mathbf{q}}=\frac{q^{\mu}q^{\nu}}{\gamma q^4}+\frac{4\pi i}{q^2}\epsilon^{\mu\nu\sigma}q_{\sigma},
\end{equation}
where $q^2=|\mathbf{q}^2|-\Omega^{2}-i0^+$ with $\Omega$ the frequency of gauge bosons.

Before proceeding, we note the analogy between Chern-Simons theory and the quantum mechanics of charged particles moving in perpendicular magnetic field. Consider the Maxwell-Chern-Simons theory in the long-wave limit, which is described by
\begin{equation}\label{eqs24}
L_{MCS}=\frac{1}{2g^2}(\partial_0A_i)^2+\frac{K}{2}\epsilon^{ij}\partial_0A_iA_j,
\end{equation}
where $K=1/2\pi$ in the current case of Chern-Simons fermionizaton.  In the limit of $g^2\rightarrow\infty$, $L_{MCS}$ reduces to a pure Chern-Simons term in the long-wave length limit. With regarding $A_i$ as the coordinate of moving particles, $g^2$ as the inverse of particle mass and $K$ the strength of magnetic field, $L_{MCS}$ then coincides with the Lagrangian describing non-relativistic particles moving in 2D plane under a perpendicular external field. This leads to Landau quantization with Landau levels spaced by cyclotron frequency as $\Omega_c=Kg^2=g^2/2\pi$. Therefore, the gauge bosons of the Maxwell theory acquire a gap, $\Omega_c$. Accordingly, the pure CS term has no excitations due to the large energy gap $\Omega_c$, however, the virtual process of absorbing and emitting bosons can affect the fermionic system via $S^{loc}_a$ in Eq\eqref{eqs9}. This suggests us integrate out the gapped local gauge field in $S^{loc}_a+S_{CS}$, i.e., Eq.\eqref{eqs9} and Eq.\eqref{eqs10}. For the microscopic processes that transfer the characteristic energy $\Omega_c$,  an induced four-fermion interaction term can be obtained in the continuum limit as
%For larger $n$, $\delta S$ contributes interaction terms with larger transfer of frequency.
%With focusing on the low-energy sector, we take into account the case where an energy quanta is transferred for absorbing/emitting one CS gauge bosons with $\Delta E=\Omega_c$.
\begin{equation}\label{eqs26}
\delta S=-\frac{\sqrt{3}t^2\Omega_c}{2}\sum_{\alpha}\int dx_0f^{\dagger}_{\mathbf{r}_0,\alpha}f_{\mathbf{r}_0,\overline{\alpha}}d^{\dagger}_{\mathbf{r}_0,\overline{\alpha}}d_{\mathbf{r}_0,\alpha}\equiv-\int dx_0 \delta H,
\end{equation}
from which one can read off the effective correction $\delta H$ to the fermion bath, namely, the Eq.(8) of the main text with $J_{eff}=\sqrt{3}t^2\Omega_c/2$. Here, the frequency of CS gauge bosons, $\Omega_c$, dominates over the energy scale of low-energy Dirac fermions for a pure CS gauge theory with $g^2\rightarrow\infty$. This is expected, because the gauge fluctuation is known to behave as a glue that mediates ``strong forces" between the fractionalized excitations.
%from which one can read off the effective correction $\delta H$ to the fermion bath, leading to Eq.(8) of the main text. Apparently, this is a gauge field-induced fermion-fermion interaction between the bath and the effective impurity.

Therefore the total effective Hamiltonian is found as $H=H^{bath}+H^{loc}_0+\delta H$, together with the constraint condition $\sum_{\alpha} d^{\dagger}_{\mathbf{r}_0,\alpha}d_{\mathbf{r}_0,\alpha}=1$, namely Eq.(5) of the main text. We note that this  single-occupation constraint can be equivalently enforced by the local impurity term, $H_{d}=\sum_{\alpha}\epsilon_dd^{\dagger}_{\mathbf{r}_0,\alpha}d_{\mathbf{r}_0,\alpha}+(U/2)\sum_{\alpha}d^{\dagger}_{\mathbf{r}_0,\alpha}d_{\mathbf{r}_0,\alpha}d^{\dagger}_{\mathbf{r}_0,\overline{\alpha}}d_{\mathbf{r}_0,\overline{\alpha}}$, with $\epsilon_d<0$ and $\epsilon_d+U>0$ and an infinite $U$.
As discussed in the main text, $\delta H$ is mapped to the  pseudospin exchange interaction, $J_{eff}$ (Eq.(9) of the main text), with the help of the constraint condition.  On the other hand, $H^{loc}_0$ in Eq.\eqref{eqs6}, along with the single-occupation condition, or equivalently $H_d$, is also mapped to the exchange interaction. This becomes obvious if we note that $H^{loc}_0+H_d$ is nothing but the Anderson impurity model, which is mapped to the exchange model via the Schrieffer-Wolf transformation. Since an infinite $U$ is required to exactly enforce  the single-occupation condition, this coupling constant of the mapped exchange model is small, and can be neglected compared to the large $J_{eff}$ induced by gauge fluctuations.

\section{Some Technical details on the gauge fluctuation and emergent pseudospin-1/2 impurity}
\subsection{Details on self-consistent calculation of fluxes}
In the main text, we have obtained the fermionized model, $H$,  which describes spinless fermions coupled to CS gauge field, i.e., Eq.(4) of the main text.   It is noteworthy that the Chern-Simons action gives no energy term of gauge fields.  The system energy is only dependent on the fluxes in the plaquettes, namely $\{\phi_{\mathbf{r}}\}\equiv\{\phi_{\mathbf{r}_1},\phi_{\mathbf{r}_2},...\}$ where $\phi_{\mathbf{r}_i}$ ($i=1,2,...N$) denotes the flux of the $\mathbf{r}_i$-th plaquette. N is the number of plaquettes on a finite-sized lattice.

For a given flux configuration  $\{\phi_{\mathbf{r}}\}$, we can obtain the fermions' minimal energy $E_{min} (\{\phi_{\mathbf{r}}\})$ by diagonalizing $H$ and summing the energies below Fermi energy.  To obtain the ground-state energy of the system, we have to vary over  $\{\phi_{\mathbf{r}}\}$, then to compare the corresponding energies $E_{min} (\{\phi_{\mathbf{r}}\}) $, and finally to find those flux configurations that can minimize $E_{min} (\{\phi_{\mathbf{r}}\})$. Thus, the ground-state energy is given by
\begin{equation}\label{eqss6add}
E_G=\mathrm{Min}[E_{min}  (\{\phi_{\mathbf{r}}\}), \{\phi_{\mathbf{r}}\}].
\end{equation}
If there is a unique flux configuration that can minimize $E_{min} (\{\phi_{\mathbf{r}}\})$ and any gauge perturbation significantly raises $E_{min} (\{\phi_{\mathbf{r}}\})$, then the system has a stable gauge flux configuration in its ground state. Otherwise, a lot of gauge flux configurations give energies $E_{min}$'s very close to $E_G$. Then, gauge fluctuations should be considered as important physical degrees of freedom in the low-energy effective theory.

We utilize the gradient descendant method to obtain $E_G$. Specifically, by starting from a generic initial flux configuration $\{\phi_{\mathbf{r}}\}=\{\phi_{0,\mathbf{r}}\}$, one can numerically calculate the gradient of $E_{min}(\{\phi_{\mathbf{r}}\})$ at the initial point $\{\phi_{0,\mathbf{r}}\}$. In the next iteration, one updates $\{\phi_{\mathbf{r}}\}$ by moving a small step, $\{d\phi_{\mathbf{r}}\}=\{d\phi_{\mathbf{r}_1},d\phi_{\mathbf{r}_2},...\}$ along the inverse direction of the gradient. This offers a locally optimized $\{\phi_{\mathbf{r}}\}$ that lowers the system energy compared to the last iteration. Then, one again evaluates  the gradient at $\{\phi_{0,\mathbf{r}}+d\phi_{\mathbf{r}}\}$ and repeats the above steps until convergence is reached. This procedure, once applied to the above model, can efficiently  produce the optimal energy and determine the fluxes self-consistently. Applying this procedure to models on ideal lattices without any defects will generate stable flux configurations that minimize the system energy. For square and honeycomb lattice, it efficiently produces the $\pi$-flux and $2\pi$-flux phase, respectively, in accordance with the Lieb's theorem.

%We now shift our attention to the fluctuations around the determined saddle point (which are negligible for stable flux phases on ideal lattices but may be essential when a strong bond defect is present). In principle, since the above procedure treats the gauge fluxes as variational parameters, it is not fully able to capture the fluctuation effect of the fluxes. However, it can still provide the numerical evidence and quantify such such  gauge fluctuations. This is demonstrated in the following.
When a strong bond defect is present, it is found that, although the $2\pi$-flux can be stabilized in most plaquettes, the flux nearby the defect, i.e., $\phi_{\mathbf{r}_0}$ cannot be fixed, even if the system energy gets converged ($10^{-7}$ accuracy in the calculations).  After optimization, it is found that $\phi_{\mathbf{r}_0}$ can take values in a region $\phi_{\mathbf{r}_0}\in[2\pi-\Delta\phi_{\mathbf{r}_0},2\pi+\Delta\phi_{\mathbf{r}_0}]$, in which the system energy becomes very close.
 To quantify such gauge fluctuations, we prepare randomly generated initial fluxes $\{\phi_{0,\mathbf{r}}\}$ and minimize the system energy based on the gradient descendant. It is found the optimized $\phi_{\mathbf{r}_0}$ relies on the initial fluxes. Therefore, it is very natural to define $\delta\phi_{\mathbf{r}_0}=(\langle\phi^2_{\mathbf{r}_0}\rangle-\langle\phi_{\mathbf{r}_0}\rangle^2)^{1/2}$, which numerically characterizes the gauge fluctuation.

\subsection{The effective pseudospin-1/2 impurity at the bond defect}
The emergence of pseudospin-1/2 impurity can be rigorously proved after combination of the Lieb's theorem and the Chern-Simons theory.

Firstly, the saddle point flux through the defect, namely through each of the two plaquettes nearby the defect at $\mathbf{r}_0$, is 0 mod $2\pi$. This has been proved both by our numerical calculation and by Lieb’s theorem.

Then, we consider the Chern-Simons theory, $L=\frac{1}{4\pi}\epsilon^{\mu\nu\rho}A_{\mu}\partial_{\nu}A_{\rho}-A_{\mu}J^{\mu}$. The equation of motion gives $J^{\mu}=\frac{1}{4\pi}\epsilon^{\mu\nu\rho}F_{\nu\rho}$. Thus, $J^0=n=\phi/2\pi$, where $\phi$ is the flux and $n$ the fermion density. This result from the continuous CS field theory is naturally cast into the following form on a lattice, i.e.,
\begin{equation}\label{eqss3}
\phi_{\mathbf{r}}=2\pi n_{\mathbf{r}}, \forall\mathbf{r},
\end{equation}
where $\mathbf{r}$ the spatial coordinate. On a lattice, we use the convention that  each plaquette is labelled by its left-bottom site, $\mathbf{r}$. Thus, the flux associated with the $\mathbf{r}$-plaquette is proportional to the fermion number at $\mathbf{r}$. The lattice site $\mathbf{r}$ consists of two sites with A and B sublattice. Therefore, the combination of the above two results directly leads to a constraint on the local fermion number. Since $\phi_{\mathbf{r}_0}=0~(\mathrm{mod}~2\pi)=2\pi n_{\mathbf{r}_0}\equiv2\pi(n_{\mathbf{r}_0,A}+n_{\mathbf{r}_0,B})$, we arrive at the integer filling at the defect site, i.e., $n_{\mathbf{r}_0,A}+n_{\mathbf{r}_0,B}=0,1,2$. Moreover, the fermionized model has particle-hole symmetry. This further fixes the filling to be exactly
\begin{equation}\label{eqss7}
n_{\mathbf{r}_0,A}+n_{\mathbf{r}_0,B}=1
\end{equation}
This local single-occupation restriction clearly generates a local two-state system. The single fermion state can occupy either A or B sublattice, thereby resulting in a pseudospin-1/2 impurity at $\mathbf{r}_0$.

 We now exactly prove Eq.\eqref{eqss3} in the CS representation on a lattice. Following from the first section in the supplemental material, the string operators, once applied to spin-exchange models on lattices, will generate phases on the bond as,
\begin{equation}\label{eqss4}
\phi_{\mathbf{r},\mathbf{r}^{\prime\prime}}=\alpha_{\mathbf{r}}-\alpha_{\mathbf{r}^{\prime}}=\int^{\mathbf{r}}_{\mathbf{r}^{\prime}}d\mathbf{l}\cdot\boldsymbol{\nabla}_{\mathbf{l}}\alpha_{\mathbf{l}}=\sum_{\mathbf{r}^{\prime}}f^{\dagger}_{\mathbf{r}^{\prime}}f_{\mathbf{r}^{\prime}}\int^{\mathbf{r}}_{\mathbf{r}^{\prime\prime}}d\mathbf{l}\cdot\mathbf{A}_{\mathbf{l}-\mathbf{r}^{\prime}},
\end{equation}
where we have defined,
\begin{equation}\label{eqss5}
\mathbf{A}_{\mathbf{l}-\mathbf{r}^{\prime}}=\boldsymbol{\nabla}_{\mathbf{l}}\mathrm{arg}(\mathbf{l}-\mathbf{r}^{\prime})=\hat{z}\times\frac{\mathbf{l}-\mathbf{r}^{\prime}}{|\mathbf{l}-\mathbf{r}^{\prime}|^2}.
\end{equation}
Then, the flux  can be calculated readily by the integral over $\mathbf{l}$ in Eq.\eqref{eqss4}, along a closed loop enclosing the plaquette, leading to
\begin{equation}\label{eqss6}
\phi_{\mathbf{r}}=\sum_{\mathbf{r}^{\prime}}f^{\dagger}_{\mathbf{r}^{\prime}}f_{\mathbf{r}^{\prime}}\oint d\mathbf{l}\cdot\mathbf{A}_{\mathbf{l}-\mathbf{r}^{\prime}},
\end{equation}
After introducing the complex coordinate $z=x+iy$, the integral  $\oint d\mathbf{l}\cdot\mathbf{A}_{\mathbf{l}-\mathbf{r}^{\prime}}$ is cast into $-i\oint dz \frac{1}{z-z^{\prime}}$, which is easily obtained by the contour integrals in the complex plane.  It generates $2\pi$ if the integral trajectories enclose $z^{\prime}$ and equals to 0 otherwise. After inserting this result into Eq.\eqref{eqss6}, Eq.\eqref{eqss3} is then proved.

%So far, two rigorous steps have been demonstrated. First, the Lieb's theorem rigorously states on honeycomb lattice that $\phi_{\mathbf{r}_0}=2\pi$ (On square lattice, since the flux configurations double the lattice unit cell, generating A and B sublattice, the Lieb's theorem also states $\phi_{\mathbf{r}_0}=2\pi$ in the enlarged unit cell). Second, the identity, $\phi_{\mathbf{r}}=2\pi n_{\mathbf{r}}, \forall\mathbf{r}$, must be satisfied, i.e., the local flux is proportional to the local fermion number.

%We have shown that the emergence of the effective two-state system can be rigorously proved after combining the Lieb’s theorem and the Chern-Simons representation. We note that this above proof is general as long as a flux phase is stabilized and the particle-hole symmetry is present for the fermionized model. This is because, for any flux phase, the ``magnetic unit" cell enjoys $2\pi$ flux. Moreover, the flux condition in Eq.\eqref{eqss3} is always satisfied in the CS representation.

%\begin{figure}[t]
%\includegraphics[width=0.55\linewidth]{FigS1.eps}
%\caption{The integral trajectories to calculate the flux within the CS fermionization formalism for honeycomb (a) and square (b) lattice.}
%\end{figure}

\section{ renormalization group analysis of the AK fixed point}
In this section, we present discussions on the RG analysis of the gauge field-induced Kondo model. Although in this model, the bare value of the parameters $V$ and $J_{eff}$ are both large, we now loosen this condition and assume tunable parameters, $V$ and $J_{eff}$, in order to sketch out the whole phase diagram. Moreover, in order to be more comprehensive, we consider a generalized ``Dirac" fermions with density of states $\rho(E)\propto |E|^{\nu}/2\pi$, where $\nu=1$ for the current model.

 Apparently, $J_{eff}=V=0$ corresponds to the local moment (LM) fixed point, where the local pseudospin is effective decoupled from the bath. Near the LM fixed point, simple power counting  directly states that both $J_{eff}$ and $V$ are irrelevant to first order. To two-loop order, $V$ bears no relevant contributions, however $J_{eff}$ receives a nontrivial renormalization, generating the following RG flow near the LM fixed point
 \begin{equation}
 \frac{d J_{eff}}{dl}=-\nu J_{eff}+J^2_{eff}-\frac{1}{2}J^3_{eff},
 \end{equation}
 and
  \begin{equation}
 \frac{d V}{dl}=-V.
 \end{equation}
It is obvious that near the LM fixed point, the local potential scattering $V$ is irrelevant and can be neglected. The situation is however different for the flow of $J_{eff}$. One observes that for $\nu<1/2$, another unstable fixed point emerges, separating the RG flows towards the LM ($J_{eff}=0$) and the strong coupling limit with $J_{eff}=\infty$. For $\nu>1/2$, which is the case here, $J_{eff}$ becomes irrelevant. Thus there is no stable Kondo fixed point with $(V,J_{eff})=(0,\infty)$ as in the usual cases, and the effective impurity always stays in the LM state in perturbative sense, as shown by Fig.4(a) of the main text. We note in passing that,  the above RG flows, derived from the anisotropic XY exchange interaction,  are of the same form as those derived from the SU(2)-invariant models. This is because the coupling term, $S^z\sigma^z$ , is automatically generated under RG flow, thus the SU(2) symmetry is always restored in low-energy.

The above analysis is based on perturbation in terms of $V$ and $J_{eff}$. Therefore, it is not reliable for the strong coupling regime far away from the LM fixed point, where the perturbation breaks down.  From the above flow equations, we know that for $\nu<1/2$,  the system is renormalized to the strong coupling limit with $J_{eff}=\infty$, when $J_{eff}$ is larger than a critical value. Besides, there is a duality between the Kondo model in the strong-coupling limit and the weak-coupling Anderson model \cite{achewsons}. This duality is naturally justified if one resorts to the 1D representation of the Wilson chain. In the strong coupling Kondo limit,  the least irrelevant operators are given by the couplings between the first site and the first several sites on the Wilson chain \cite{qlli}, which physically describe the residual hybridization between the formed Kondo singlet and the bath fermions \cite{achewsons}.   Such observation suggests us investigate the  dual Anderson impurity model with a weak Hubbard interaction, $U\rightarrow0$. This dual model captures the physics in the strong coupling regime ($J_{eff}=\infty$). The dual Anderson model is given by
 \begin{equation}
 H_{ad}=\sum_{\sigma}\int^{\infty}_{-\infty}d\epsilon\epsilon f^{\dagger}_{\epsilon,\sigma}f_{\epsilon,\sigma}+\epsilon_d\sum_{\sigma}d^{\dagger}_{\sigma}d_{\sigma}+g\int d\epsilon(f^{\dagger}_{\epsilon,\sigma}d_{\sigma}+h.c.),
 \end{equation}
 where the $d$-fermions here denotes the local orbitals describing the first site of the Wilson chain in the strong coupling limit. It mainly describes the degrees of freedom of the formed Kondo singlet. $g$ here denotes the hybridization between the singlet and the bath fermions.

We note that, the original potential scattering $V$, which is a measure of the local particle-hole asymmetry, becomes implicit in the above Anderson model. Since only for $\epsilon_d=-U/2$ is the particle-hole symmetry preserved, the effect of $V$ is now captured by assuming a general nonzero $\epsilon_d$ with $\epsilon_d\neq -U/2$. $\epsilon_d$ constitutes one of the running parameters under RG flow.  We perform the standard  RG analysis of the above weakly-coupling asymmetric Anderson model. The following RG flow is derived as (for $\nu=1$),
    \begin{equation}
 \frac{d g}{dl}=-\frac{3}{2}g^3,
 \end{equation}
 and
     \begin{equation}
 \frac{d \epsilon_d}{dl}=\epsilon_d+g^2-3g^2\epsilon_d.
 \end{equation}
 From above, we know that, besides the LM fixed point $(\epsilon_f,g)=(-\infty,0)$, a new fixed point is generated as $(\epsilon_f,g)=(\infty,0)$. $\epsilon_f=\infty$ suggests that the doublet state is frozen due to the large energy penalty. Moreover, $g=0$ means that the singlet is barely coupled to the bath at this fixed point, generating a stable Kondo singlet where the effective pseudospin  of the impurity becomes fully screened, as shown by Fig.1(b) of the main text. In addition, at this fixed point,  $\epsilon_d\neq -U/2$. This reflects the fact that the particle-hole symmetry is violated by the local potential  $V$, a key feature of the AK fixed point.

\section{some details on the NRG calculations}
In this section, we present some details on the NRG calculation. We study the mapped Kondo model by the full-density-matrix (FDM) NRG. The density of states of the non-interacting bath takes a linear form,
\begin{equation}
\rho(\epsilon)=\frac{|\epsilon|}{2\pi},
\end{equation}
where, $\epsilon\in[-1, 1]$. The bath is discretized logarithmically and mapped to a semi-infinite Wilson chain with exponentially decaying hoppings, and the impurity coupled to the first chain site via the Kondo constant $J_{eff}$. The potential scattering $V$ is correspondingly mapped to a local on-site energy $V$ at the boundary of the Wilson chain.  The chain is diagonalized iteratively while discarding high-energy states, thereby zooming in on low-energy properties: the finite-size level spacing of a chain ending at site $k$ is of order $\omega_k\propto \Lambda^{-k/2}$. Here $\Lambda>1$ is a discretization parameter, chosen to be 2 in this work. We use the full-density-matrix NRG~\cite{Andreasa} method that is implemented in the QSpace~\cite{Andreasb} tensor library to solve this model. We keep 1000 multiplets in the diagonalizations. The imaginary part of the impurity dynamical susceptibility, $\chi(\omega)=-\text{Im}\langle S||S\rangle_{\omega}$, was calculated at temperature $T=10^{-10}$.  For a Fermi liquid, $\chi(\omega)\propto \omega$ in low energies.

\section{Thermal conductivity and possible generalizations}
\subsection{Thermal conductivity}
We now present in this section some technical details for calculation of the thermal conductivity at the AK fixed point. Our calculation is based on the conformal field theory (CFT) combined with linear response theory. We first map the emergent Kondo problem, i.e.,  Eq.(7) and (9) in the manuscript, into a 1D model. We utilize the rotation symmetry with respect to the impurity, and make transformation to the orbital angular momentum (OAM) space, generating relevant partial waves labelled by the OAM, $L_z$. Next, we resort to the eigen-basis of the total angular momentum $J_z=L_z+\sigma_z/2$, where $\sigma_z$ is the pseudospin-1/2 as defined in the main text.  These steps lead to the 1D model as,
\begin{equation}\label{eqssd10}
H^{bath}=\sum_{m=1,2}\sum_{\sigma}\int^{\infty}_{-\infty} d\epsilon(\epsilon-\mu)d^{\dagger}_{\epsilon,m,\sigma}d_{\epsilon,m,\sigma},
\end{equation}
and
\begin{equation}\label{eqssd11}
 H_K=g_1\int^{\infty}_{-\infty}d\epsilon d\epsilon^{\prime}d^{\dagger}_{\epsilon,1,\sigma}\boldsymbol{\sigma}_{\sigma\sigma^{\prime}}\cdot\mathbf{S}_{p}d_{\epsilon^{\prime},1,\sigma^{\prime}}
 +g_2\int^{\infty}_{-\infty}d\epsilon d\epsilon^{\prime}d^{\dagger}_{\epsilon,2,\sigma}\boldsymbol{\sigma}_{\sigma\sigma^{\prime}}\cdot\mathbf{S}_{p}d_{\epsilon^{\prime},2,\sigma^{\prime}},
\end{equation}
where $d_{\sigma,m\sigma}$ is the transformed fields describing the bath fermions, and one should not confuse them with the impurity states defined in the manuscript. Since the effect of $V$, as shown by the manuscript, is to introduce an effective  local chemical potential, we introduce $\mu$ in the bath while neglect the $V$ term. This facilitates the following calculations. Moreover, it does not generate any deviations as long as the low-energy fixed points are considered, which has been verified by NRG calculations.  $m=1,2$ is the ``channel" flavor, and $g_1=(g_d+g_t)$, $g_2=(g_d-g_t)$, with $g_d$ and $g_t$ the intra- and inter-valley scattering. For the case discussed in the manuscript, $g_d=g_t=J_{eff}$, thus $g_2=0$, leading only to a single-channel Kondo model with Fermi liquid (FL) behavior. By contrast, for $g_t=0$, i.e., when the inter-valley scattering is suppressed, the two-channel Kondo physics will take place, giving rise to possible non-Fermi liquid (NFL) behavior.

Based on Eq.\eqref{eqssd10} and Eq.\eqref{eqssd11}, we now calculate the thermal conductivity based on the results from CFT and the linear response theory. In the CFT formalism, the above 1D model can be further mapped to half-infinite chain with left and right movers denoted by fields $d_{L/R,m,\sigma}$ \cite{Afflecka}. We consider the single-particle Green's function defined by $\langle d^{\dagger}_{L,m,\sigma}(z_1)d_{R,m,\sigma}(z_2)\rangle$, where $z$ lies in the complex plane representing for the 1+1D spacetime. We obtain $\langle d^{\dagger}_{L,m,\sigma}(z_1)d_{R,m,\sigma}(z_2)\rangle=0$ for the case with no boundary, and $\langle d^{\dagger}_{L,m,\sigma}(z_1)d_{R,m,\sigma}(z_2)\rangle_{Free}=1/(z_1-\overline{z}_2)$ for a trivial boundary (corresponding to the weak coupling regime). The correlation function in the strong-coupling Kondo regime can be calculated via the boundary state that in turn obtained by fusion \cite{Afflecka,Affleckb,Affleckc,Ludwig,Affleckd,Afflecke,Ludwigf}, leading to the scattering matrix $S$ connecting the correlations as $\langle d^{\dagger}_{L}(z_1)d_R(z_2)\rangle_{Kondo}=S\langle d^{\dagger}_{L,m,\sigma}(z_1)d_{R,m,\sigma}(z_2)\rangle_{Free}$ , where  $S=-1$ and $S=0$ for the FL and NFL fixed point, respectively.
Assuming a dilute impurity with density $n_{imp}$, the scattering time $\tau^{-1}_s=-2\mathrm{Im}\Sigma^R(\omega)$, where $\Sigma^R(\omega)$ is the retarded self-energy that is related to $S$ \cite{Afflecke}.

On the other hand, the thermal current can be readily derived from Eq.\eqref{eqssd10} and Eq.\eqref{eqssd11} as,
\begin{equation}\label{eqsn12}
\mathbf{j}_E=-\hat{k}\sum_{m,\sigma}\int d\epsilon (\epsilon-\mu)d^{\dagger}_{\epsilon,m\sigma}d_{\epsilon,m,\sigma},
\end{equation}
where $\hat{k}=\mathbf{k}/k$. The thermal conductivity can be evaluated through $T\sigma_E(\omega)=-\lim_{\omega\rightarrow0}\mathrm{Im}\pi^R(\omega)/\omega$, where $\pi^R$ is the retarded current-current correlation function. We firstly calculate the current correlation function in Matsubara form as,
\begin{equation}\label{eqsn13}
\pi(i\omega_n)=-\frac{1}{3}\int^{\beta}_0 d\tau e^{i\omega_n\tau} \langle \hat{T}_{\tau}\mathbf{j}_E(\tau)\cdot\mathbf{j}_E(0)\rangle,
\end{equation}
where $\hat{T}$ is the imaginary time ordering. Inserting Eq.\eqref{eqsn12} into Eq.\eqref{eqsn13}, we obtain
\begin{equation}\label{eqsn14}
\pi(i\omega_n)=\frac{1}{3}\int d\epsilon\xi^2\frac{1}{\beta}\sum_{i\nu_n} g(\epsilon,i\omega_n+i\nu_n)g(\epsilon,i\nu_n),
\end{equation}
where $\xi=\epsilon-\mu$, and $g(\epsilon,i\omega_n)$ is the Matusbara Green's function of the d-fields.  Performing the sum of Matusbara frequency $i\nu_n$ and after insertion of $\pi(i\omega_n)$ into $\sigma_E$, one obtains that
\begin{equation}\label{eqsn15}
T\sigma_E=\frac{1}{3}\int d\epsilon\xi^2\int d\nu\delta(\epsilon-\nu)\tau_s(-\frac{\partial}{\partial{\nu}}n_F(\nu)),
\end{equation}
where $n_F(\nu)=1/\mathrm{exp}[\beta(\nu-\mu)+1]$ is the Fermi distribution function.  We consider lowest-order $T$ behavior of $T\sigma_E$. Since $\tau_s$ is dependent on $S$ which further relies on higher-order $T$-terms, its $T$-dependence can be neglected for low $T$. After inserting the Fermi distribution function and completing the integrals, it is straightforward to find that the right-hand-side of Eq.\eqref{eqsn15} is proportional to $T^2$. Thus, we obtain the thermal conductivity at low temperature as,
\begin{equation}\label{eqssd16}
\sigma_E(T)/T=\pi^3\rho_0/[9(1-S)n_{imp}],
\end{equation}
where we assume a dilute distribution of impurities with density $n_{imp}$ and $\rho_0$ is the effective density of states at Fermi energy, which is effectively enhanced by $V$ and can be non-vanishing at the Dirac point.  At finite temperatures, the higher-order corrections from the irrelevant operators in CFT come into play  \cite{Ludwigf,Afflecke}. Following the same procedures above, these operators generate different scaling behaviors for the FL and NFL fixed point, which are respectively obtained as $\sigma^{FL}_E(T)/T=\pi^3\rho_0/18n_{imp}+aT^2$ and $\sigma^{NFL}_E(T)/T=\pi^3\rho_0/9n_{imp}+bT^{1/2}$, where $a$, $b$ are universal coefficients. Therefore, for the more relevant case in the manuscript with $g_d=g_t=J_{eff}$, we predict the FL Kondo fixed point, and it displays  the $\sim T^2$ scaling of thermal conductivity at low-temperatures.

\subsection{Possible generalizations}
Our predictions in this work specifically include: (a) the enhancement of gauge fluctuations by bond defect, (b) the gauge-fluctuation-induced local fermion-fermion interactions and (c) emergent Kondo-like local resonance.
\begin{figure}[t]
\includegraphics[width=0.85\linewidth]{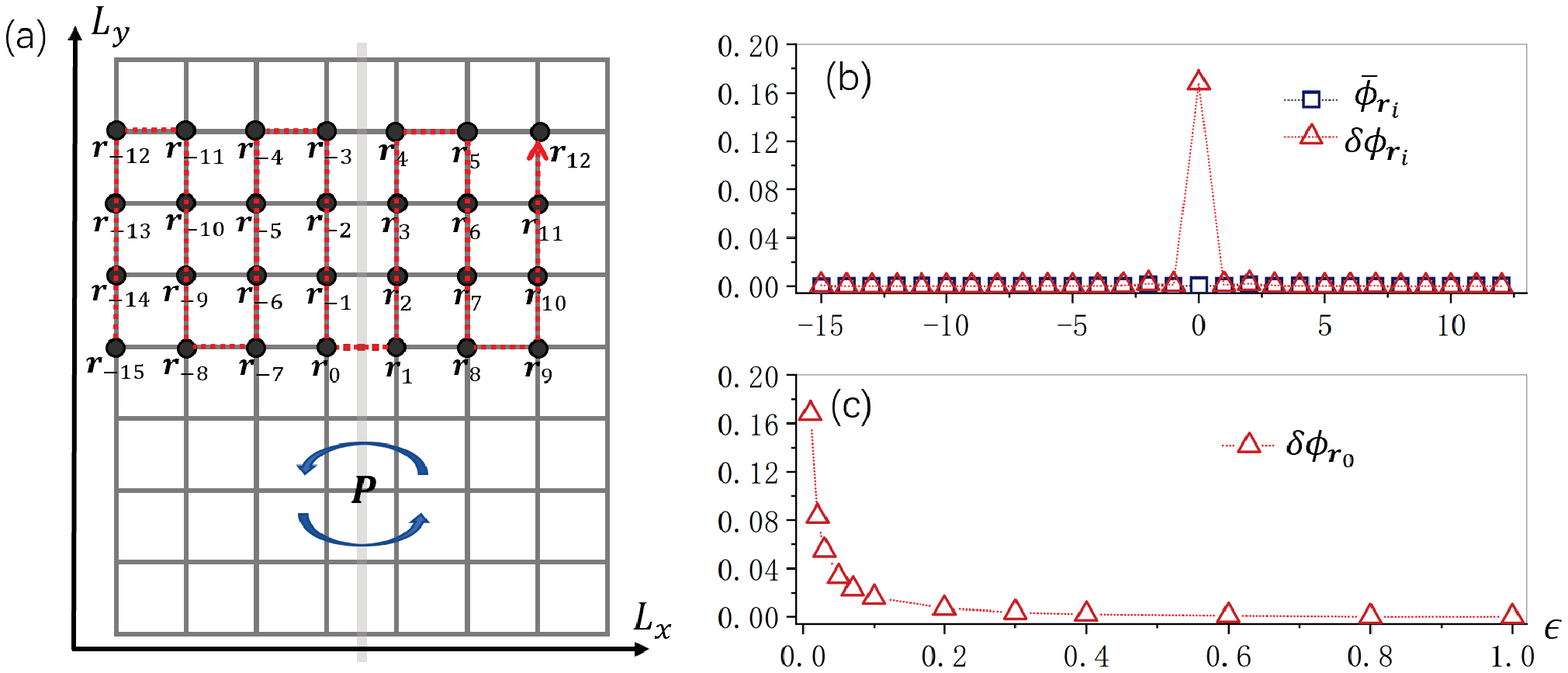}
\caption{The effect of the bond defect on the flux phase on square lattice.   (a) The quenched bond is placed at the center of a $8\times9$ lattice. The system respects the reflection symmetry $\mathbf{P}$. (b) The calculated $\overline{\phi}_{\mathbf{r}}$ and  $\delta\phi_{\mathbf{r}}$, which are defined as $\overline{\phi}_{\mathbf{r}_i}=\langle\phi_{\mathbf{r}_i}\rangle-\pi$, and $\delta\phi_{\mathbf{r}_i}=(\langle\phi^2_{\mathbf{r}_i}\rangle-\langle\phi^2_{\mathbf{r}_i}\rangle^2)^{1/2}$, with the $\langle...\rangle$ denoting the average over the random initial states (obtained with 200 random initial configurations). The x-axis denotes the 28 plaquettes along the dashed trajectory in (a), from the $\mathbf{r}_{-15}$ to $\mathbf{r}_{12}$. The $\mathbf{r}_0$-plaquette is the one adjacent to the quenched bond, whose hopping is taken as $\epsilon=0.01$. (c) $\delta \phi_{\mathbf{r}_0}$ as a function of  $\epsilon$. The n.n. hopping is taken as $t=1$.}
\end{figure}

The first point in (a) is illustrated by Fig.1(a) of manuscript. This understanding is general and not  restricted to any specific lattice. Generally, we can consider a generic flux state on different lattices. Once the system energy is minimized, the flux in all the closed loops formed by the lattice bonds can be determined, using, for example, the gradient descendant method illustrated in the above section. In this case, if a bond is removed, the plaquettes that share this bond will be connected. This will generally allow gauge fluctuations in each plaquette. Therefore, (a) is expected to be a general observation. To further show this more explicitly, we performed additional calculations on square lattice. As shown by Fig.S1, we observe that, despite that the saddle point values of fluxes now converge to $\pi$, we do observe that the bond defect significantly enhances the local gauge fluctuations, $\delta\phi_{\mathbf{r}_0}$, which is non-negligible for a strong bond defect with  $\epsilon\ll t$. This clearly justifies the lattice-independence of this observation.

The second point in (b) is obtained by a general method, namely by integrating out the local Chern-Simons gauge field. It is known that the gauge field can meditate local fermion-fermion interactions. Such a general procedure will lead to similar local interactions in other models as well.  The third point in (c) follows directly from (b) and the formation of a local two-state system. The latter condition is general as long as the flux phase is stable and the particle-hole symmetry is present. The logic to prove the emergence of the two-state system as been discussed in the previous section.
Then, following our technical steps, an effective model mimicking the spin-1/2 Kondo exchange term will generally be generated. Moreover, the Kondo exchange, as we know, is relevant in renormalization group sense, and will favor the Kondo behavior in low-energies, as long as the fermion bath is gapless (if there is gap, the competition between the gap and the Kondo scale should take place, which complicates the situation).

Another interesting direction for further exploration is to consider possible generalizations to other systems with different types of excitations. Regarding to this aspect, we would like to make the following comments, in the hope of stimulating further investigations.  First, it should be noted that the spin-1/2 systems (the starting point of this work) can be equivalently described by hardcore bosons. Therefore, it is possible to realize the flux phases or other spin liquids in boson systems with strong local boson-boson interaction. Second, we point out that it will be very important to focus on those bosonic systems, whose single-particle Hamiltonians display exotic energy bands, especially those with large energy degeneracies, such as the flat band and moat band. For these systems, it is likely that the bosons will be further “fractionalized”, leading to fermionic partons and emergent gauge fields. Therefore, we expect that addition bond defects could have nontrivial effects in these exotic bosonic systems, which may in turn be realized experimentally, for example, in the dilute atomic systems.

%Figure Captions \newline

\end{document}